\documentstyle[emulateapj,danonecolfloat]{article}
\input psfig.sty

\def\spose#1{\hbox to 0pt{#1\hss}}
\def\simlt{\mathrel{\spose{\lower 3pt\hbox{$\mathchar"218$}}
     \raise 2.0pt\hbox{$\mathchar"13C$}}}
\def\simgt{\mathrel{\spose{\lower 3pt\hbox{$\mathchar"218$}}
     \raise 2.0pt\hbox{$\mathchar"13E$}}}
\def\simpropto{\mathrel{\spose{\lower 3pt\hbox{$\mathchar"218$}}
     \raise 2.0pt\hbox{$\propto$}}}
\newcommand\lsim{\mathrel{\rlap{\lower4pt\hbox{\hskip1pt$\sim$}}
        \raise1pt\hbox{$<$}}}
\newcommand\gsim{\mathrel{\rlap{\lower4pt\hbox{\hskip1pt$\sim$}}
        \raise1pt\hbox{$>$}}}

\def\ie{{\it i.e. }}
\def\eg{{\it e.g. }}
\def\rmb{{\rm b}}
\def\rmf{{\rm f}}
\def\rmg{{\rm g}}
\def\rmm{{\rm m}}
\def\rmp{{\rm p}}
\def\bfC{{\bf C}}
\def\bfd{{\bf d}}

\def\chieff{{\chi_{\rm eff}}}

\begin{document}
\twocolumn[%%% Begin front material

%%%%%%%%%%%%%%%%%%%%%%%%%%%%%

%\tighten
%\eqsecnum
%\received{4 August 1988}
%\accepted{23 September 1988}
%\journalid{337}{15 January 1989}
\articleid{11}{14}

\submitted{\today}
%\submitted{\today. To be submitted to ApJ.}
%\submitted{Submitted to ApJL September 16; accepted February 2}

\title{Isolating Geometry in Weak Lensing Measurements}
\author{Jun Zhang$^{1,2}$, Lam Hui$^{1,2,3}$ and Albert Stebbins$^{1,3}$}
\affil{$^1$ Theoretical Astrophysics, 
Fermi National Accelerator Laboratory, Batavia, IL 60510\\
$^2$ Department of Physics, Columbia University,
New York, NY 10027\\
$^3$ Department of Astronomy and Astrophysics, University of 
Chicago, IL 60637\\
{\tt jz203@columbia.edu, lhui@fnal.gov, stebbins@fnal.gov}}
%\altaffiltext{1}{{\tt jz203@columbia.edu}}
%\altaffiltext{2}{{\tt lhui@astro.columbia.edu}}
                                          
\begin{abstract}
Given a foreground galaxy-density field or shear field, its cross-correlation
with the shear field from a background population of source galaxies scales
with the source redshift in a way that is specific to lensing.  Such a
source-scaling can be exploited to effectively measure geometrical
distances as a function of redshift and thereby constrain dark energy
properties, free of any assumptions about the 
galaxy-mass/mass power spectrum (its shape, amplitude or growth).
Such a geometrical method can yield a
$\sim 0.03 - 0.07 f_{\rm sky}^{-1/2}$ 
measurement on the dark energy abundance and equation of state, for a photometric
redshift accuracy of 
$\Delta z \sim 0.01 - 0.05$ and a survey
with median redshift of $\sim 1$.
While these constraints are weaker than conventional weak lensing methods, 
they provide an important consistency check because 
the geometrical method carries less theoretical baggage: 
there is no need to assume any structure formation model ({\it e.g.} CDM). 
The geometrical method is at the most conservative end of a whole
spectrum of methods which obtain smaller errorbars by making more
restrictive assumptions -- we discuss some examples.
Our geometrical approach differs from previous investigations along similar lines
in three respects. First, the source-scaling we propose to use is 
less demanding on the photometric redshift accuracy. Second, the scaling works for
both galaxy-shear and shear-shear correlations.
Third, we find that previous studies underestimate the statistical errors
associated with similar geometrical methods, the origin of which
is discussed.
\end{abstract}


]%%% End front material

\section{Introduction}
\label{intro}

Weak gravitational lensing has emerged to become an important probe of
cosmology
(\eg \cite{wittman00,vW00,bacon00,kwl00,maoli01,rhodes01,h02,jarvis,pen03};
see review by \cite{bs01}).  Much of the current discussion on potential dark
energy constraints from weak lensing has focused on the use of the
shear/convergence power spectrum, or equivalent measures, as a function of
source redshift (\eg \cite{hu02,aba03,bv03} but see \eg
\cite{hui99,vw01,bb01,huterer02,munshi03,alex03,tj03} for dark energy constraints from
skewness or bispectrum).  In these types of investigations, information about dark energy
(both its abundance and equation of state) is encoded in the combination of
geometrical distances and fluctuation growth rate that determines the observed
lensing power spectrum.  In this paper, we would like to pose and answer the
question: is it possible to separate out the information purely from geometry
i.e. irrespective of details of the mass power spectrum and its growth ?  Such
an exercise is useful because a method to do so allows us to derive dark energy
constraints without making assumptions about the underlying large scale
structure model (\eg Cold Dark Matter, Gaussian initial conditions, etc).
Comparing lensing constraints obtained via such a geometrical method against 
lensing constraints that carry more theoretical baggage provides an important consistency check.
Moreover, a geometrical method allows us to make use of 
lensing measurements on small scales, scales which are often ignored in
conventional methods because of worries about the ability to predict
the nonlinear power spectrum accurately.

Our discussion is divided as follows. In \S\ref{scaling}, we point out an
interesting scaling of lensing signals (i.e.  shear-shear and
galaxy-shear power spectra) with the source distance. Such a scaling
can be used to obtain
essentially an estimate of angular diameter distance (or more precisely, combinations of 
angular diameter distances) as a function of source redshift, 
without making any assumptions about the mass/galaxy power spectrum. 
We contrast this scaling with a different interesting scaling investigated 
by Jain \& Taylor (2003),
especially in terms of the demand on photometric redshift accuracy.
The scaling we focus on can be applied
to both galaxy-shear and shear-shear data, whereas the scaling of Jain \& Taylor
applies only to galaxy-shear.
To understand what kind of constraints one could obtain about dark energy from
our geometrical method, we perform a Fisher matrix analysis in \S \ref{fisher}.
The conclusions are summarized in Fig. \ref{OmegaVw_20} and \ref{Omegaww_prime_20}.
The geometrical method above is a very conservative one: it makes
absolutely no assumptions about the underlying large scale structure and its evolution.
In \S\ref{2ws} we investigate a method at the other end 
of the spectrum: it assumes the shape of the mass/galaxy power
spectrum is known. It differs from more conventional methods (such as
lensing tomography of Hu 2002, Abazajian \& Dodelson 2003) only
in that the geometrical information and growth rate information are
separated to provide a consistency check. 
In practice, there is of course a whole continuum of methods to obtain
dark energy constraints from lensing data, varying
from the most conservative (like the geometrical method emphasized here)
to ones that make strong large scale structure assumptions.
We conclude in \S\ref{discuss}.

A word on the history of this project is in order.  When we started, our
initial focus was on shear-shear
correlation. Since then, an elegant paper by Jain and Taylor (2003, \cite{jt03} 
hereafter) appeared which addressed similar issues, but using the galaxy-shear
correlation (see also interesting developments in 
Bernstein \& Jain 2003 [\cite{bj03}]).  Therefore we
decide to include both in our discussion here.  While our results are in
qualitative agreement, we find quantitative differences.  In
particular we find dark energy constraints that are weaker than \cite{jt03}. 
As we will explain in detail later, it is not {\it a priori} obvious whose constraints
should be stronger. This is because we focus on a source-scaling of lensing signals
that is different
from JT03. Our scaling is less demanding on the photometric redshift accuracy and
can be applied to both galaxy-shear and shear-shear correlation data, but introduces
more free parameters.
{\it However}, even if we
employ exactly the same scaling adopted by \cite{jt03}, we find statistical
errors that are larger than JT03, the origin of which is discussed in detail in Appendix A.
Related issues are discussed by Song \& Knox (2003) and Hu \& Jain (2003).

A word on terminology. \cite{jt03} used the term cross-correlation tomography to
describe their method. This term can also be used to refer to the technique
of cross-correlating shear/galaxy-density fields from different redshifts in general
(e.g. \cite{tw03}). What we would like to focus on, as in the case of \cite{jt03},
is the use of the cross-correlation technique to extract cosmological constraints that
are purely geometrical in origin. To avoid confusion, we will generally not
use the term cross-correlation tomography. We will simply refer to our approach as
a geometrical method. To distinguish the source-scaling we exploit from the one used by \cite{jt03},
we refer to ours as the offset-linear scaling (as opposed to the linear scaling adopted by \cite{jt03}).
The difference between these two scalings will be explained in the next section.

In the bulk of this paper,
the term shear is loosely used to refer to its scalar part (i.e. the convergence).
For simplicity, most of our expressions focus on correlations involving
the convergence, and they assume a flat universe.
Expressions for the more directly observable components of shear 
and for a non-flat universe are given in Appendix B. 

\section{A Useful Scaling of Lensing Signals with Source Distance}
\label{scaling}

Here we consider several redshift distributions of galaxies some of which are
considered foreground distribution (labeled by $f$) and other considered
background galaxies (labeled by $b$).  The idea being that the background
galaxies are behind the foreground galaxies.  One measures the lensing shear
field of the different background galaxy populations and correlates it with
either the shear field or the surface number density
fluctuations of the foreground galaxies.  By determining how these correlations
scale with the redshift distribution of the background galaxies we hope to
learn about the cosmology in a way which is independent of assumptions about
inhomogeneities in the universe and depends only on the overall geometry.

It will be important that the foreground galaxies are indeed in front of the
background galaxies and hence that they do not overlap in redshift with the
background galaxies.  
\footnote{However, it is completely unimportant whether different background (or foreground)
populations overlap, or even whether they contain common
members.}
One can measure precisely using spectroscopy or estimate approximately
using multi-color photometry the redshift distribution of the different
populations.   With spectroscopic redshifts it is a simple matter to assure
that the foreground distribution and the background distribution 
overlap very little while with photometric redshifts this
requires more care.  In \S\ref{fisher} below we will show how a small
contamination of background galaxies in front of foreground galaxies affects
the results.

In this paper we find it more useful to express everything in terms of comoving
distance from the observer, $\chi$, rather than redshift $z$. Of course,
observationally one measures $z$ and can only infer exact values of $\chi$
once one {\it assumes} a cosmology, which then gives you the function
$\chi(z)$. The idea is to find cosmological parameters that give distances
which best match the observed lensing correlations. 
Of crucial importance will be the $z$-distributions of
galaxy populations, $dN(z)/dz$, but below we will use the distance
distribution
%\footnote{One may  not want to weight all galaxies equally when
%computing a correlation and the distributions used in practice would also be
%weighted, but this complication is not important for this paper.}
\begin{equation}
\label{DistanceDistribution}
W(\chi(z))\equiv{{dN(z)\over dz}\over
 {d\chi(z)\over dz}\,\int_0^\infty dz'\,{dN(z')\over dz'}}
\end{equation}
so that $\int d\chi\,W(\chi)=1$.  We will add a $f$ or $b$ subscript to $W$
for foreground or background populations, respectively.  Other cosmological
quantities we will use are the scale factor $a(\chi)$ defined by
$a(\chi(z))=1/(1+z)$, the Hubble parameter $H(\chi)=-c a'(\chi)/a(\chi)^2$, and
$\Omega_{\rmm0}$ is the present density of matter (dark + baryonic) in units of
the critical density.  We define $H_0\equiv H(0)$ and $c$ is the speed of
light.

For simplicity we assume a flat universe in the bulk of the paper. All
expressions, in particular the scaling of interest, can be generalized to a
non-flat universe as discussed in Appendix B. Also, the expressions in the bulk
of the paper are given in Fourier space. The real space counterparts are
discussed in Appendix B as well.

%AJS3 this equation is never used, although it is useful background material
%The lensing convergence measured from some background population of source
%galaxies is basically a projected form of mass fluctuation:
%\begin{eqnarray}
%\label{kappa}
%&&\kappa(\vtheta)={3 \Omega_{\rmm0} H_0^2 \over 2 c^2} \int_0^\infty 
%{d\chi_\rmf \over a(\chi_\rmf)} \\ \nonumber
%&&\hskip10pt\times\int_{\chi_\rmf}^\infty d\chi_\rmb\,W_b(\chi_\rmb)\, 
%{\chi_\rmf(\chi_\rmb-\chi_\rmf)\over\chi_\rmb}\,
% \delta(\vtheta, \chi_\rmf) 
%\end{eqnarray}

We are interested in 2 kinds of correlations. One is correlating the background
shear ($\gamma$) field with the foreground galaxy density field, and the other
is with some foreground $\gamma$ field.  The first is usually referred to as
{\it galaxy-galaxy lensing} and the second is known as {\it shear-shear
correlation}.  In both cases the shear that is correlated is only the scalar
(a.k.a. G-mode or E-mode) component of the shear pattern
\footnote{Also known
as ``electric-'', or ``gradient-'' component.  This excludes the other
component, known as ``curl-'', ``C-'', ``magnetic-'', ``B-', or pseudoscalar-
component, which will have a much smaller signal and less useful information
for our purposes.} (see \cite{stebbins96}). Unless otherwise stated, we will use 
$\gamma$ to refer to this scalar part: the convergence. 
Using a Limber approximation for
small angles (large $\ell$) the resulting angular cross power spectra,
$P_{\rmg\gamma}(\ell)$ and $P_{\gamma\gamma}(\ell)$ can be written as
(Blandford et al. 1991, Miralda-Escude 1991, \cite{kaiser92}, Jain \& Seljak
1997) 
\begin{eqnarray}
\label{Pgs}
P_{\rmg\gamma}(\ell;f,b)={3\Omega_{\rmm0} H_0^2 \over 2 c^2}
\int {d\chi_\rmf\over a(\chi_\rmf)} W_f(\chi_\rmf)
\int d\chi_\rmb W_b(\chi_\rmb) \nonumber \\ 
\times{\chi_\rmb-\chi_\rmf\over\chi_\rmb\chi_\rmf}\,
P_{\rmg\delta}(\frac{\ell}{\chi_\rmf},\chi_\rmf)\,\Theta(\chi_\rmb-\chi_\rmf)
\end{eqnarray}
and
\begin{eqnarray}
\label{Pss}
&&\hskip-10pt 
P_{\gamma\gamma}(\ell;f;b)=\left({3\Omega_{\rmm0}H_0^2\over 2 c^2}\right)^2 \\
\nonumber && \times \int d\chi_\rmf W_f(\chi_\rmf) 
\int d\chi_\rmb W_b(\chi_\rmb) \\ \nonumber
&&\hskip-15pt\times\int {d\chi \over a(\chi)^2}
{\chi_\rmb - \chi \over \chi_\rmb}
{\chi_\rmf - \chi \over \chi_\rmf} P_{\delta\delta}(\frac{\ell}{\chi},\chi)
\Theta(\chi_\rmb-\chi)\,\Theta(\chi_\rmf-\chi)
.\end{eqnarray}
Here $\Theta(\chi_\rmb-\chi)$ is the Lorentz-Heaviside function which is unity
if $\chi<\chi_\rmb$, and zero otherwise.  Also $P_{g\delta} (k,\chi)$ and
$P_{\delta\delta}(k,\chi)$ are respectively the 3-d galaxy-mass power spectrum
and 3-d mass power spectrum, both evaluated at 3-d wavenumber $k$ and at a time
corresponding to distance $\chi$, and $\ell$ is the angular wavenumber.  As
always with the Limber approximation there is a one-to-one correspondence with
the 3-d wavenumber and angular wavenumber at a given distance $\chi$:
$k\leftrightarrow\frac{\ell}{\chi}$.

\subsection{Offset-Linear Scaling}

The key step for the purpose of this paper is to note that if the foreground
distribution $W_f$ and the background distribution $W_b$ overlap very
little then it is an excellent approximation to make the substitution
\begin{equation}
\label{GalGalApprox}
W_f(\chi_\rmf)\,W_b(\chi_\rmb)\,\Theta(\chi_\rmb-\chi_\rmf)\rightarrow
W_f(\chi_\rmf)\,W_b(\chi_\rmb)
\end{equation}
in eq.~[\ref{Pgs}], or
\begin{eqnarray}
\label{ShearApprox}
&&W_f(\chi_\rmf)\,W_b(\chi_\rmb)\,\Theta(\chi_\rmb-\chi)\,
\Theta(\chi_\rmf-\chi)
\\ \nonumber
&&\hskip35pt\rightarrow W_f(\chi_\rmf)\,W_b(\chi_\rmb)\,\Theta(\chi_\rmf-\chi)
.\end{eqnarray}
in eq.[\ref{Pss}].  Under this approximation the angular power spectrum will
exhibit an {\it offset-linear scaling}:
\begin{eqnarray}
\label{approxscaling}
P_{\rmg\gamma}  (\ell;f,b)\approx F(\ell;f)+G(\ell;f)/\chieff(b) 
\\ \nonumber
P_{\gamma\gamma}(\ell;f,b)\approx A(\ell;f)+B(\ell;f)/\chieff(b)
\end{eqnarray}
where
\begin{eqnarray}
\label{chieff}
{1\over \chieff (b)}\equiv\int d\chi_\rmb 
W_b(\chi_\rmb)\,{1\over \chi_\rmb} 
\end{eqnarray}
and
\begin{eqnarray}
\label{FGAB}
&& 
F(\ell;f)\equiv{3\Omega_{\rmm0}H_0^2\over2c^2}
\int {d\chi_\rmf\over a(\chi_\rmf)}
\,{W_f(\chi_{\rmf}) \over \chi_\rmf} P_{g\delta}(\frac{\ell}{\chi_\rmf}, 
\chi_{\rmf}) \\
\nonumber && 
G(\ell; f) \equiv - {3\Omega_{\rmm0} H_0^2 \over 2 c^2} 
\int{d\chi_\rmf\over a(\chi_\rmf)}\,W_f(\chi_\rmf)\,
P_{g\delta}(\frac{\ell}{\chi_\rmf}, \chi_{\rmf}) \\
\nonumber &&
A(\ell;f)\equiv\left({3\Omega_{\rmm0} H_0^2 \over 2 c^2}\right)^2
\int {d\chi_\rmf}\,W_f(\chi_\rmf)\\
&&\nonumber
\hskip75pt\times\int_0^{\chi_\rmf}{d\chi \over a(\chi)^2}\,
{\chi_\rmf-\chi\over\chi_\rmf}\,P_{\delta\delta}(\frac{\ell}{\chi}, 
\chi) \\
\nonumber &&
B(\ell;f)\equiv-\left({3\Omega_{\rmm0} H_0^2 \over 2 c^2}\right)^2
\int {d\chi_\rmf}\,W_f(\chi_\rmf)
\\ \nonumber 
&&\hskip70pt\times\int_0^{\chi_\rmf}{d\chi\over a(\chi)^2}\,
{\chi_\rmf - \chi \over \chi_\rmf}\,\chi\,P_{\delta\delta}(\frac{\ell}{\chi},
\chi)
\end{eqnarray}
%\begin{eqnarray}
%&& H(\ell; i,j) \equiv - {3\Omega_{\rmm0} H_0^2 \over 2 c^2}
%\int {d\chi\over a} W_i (\chi) 
%\int d \chi' W_j (\chi') \\ \nonumber
%&& \quad \quad \quad \quad 
%{\chi' - \chi \over \chi' \chi} P_{g\delta} (k = \ell/\chi) 
%S (\chi' < \chi)
%\end{eqnarray}
%\begin{eqnarray}
%&& D(\ell; i,j) \equiv - \left({3\Omega_{\rmm0} H_0^2 \over 2 c^2}\right)^2
%\int {d\chi''} W_i (\chi'') \int d \chi' W_j (\chi')
%\\ \nonumber 
%&& \int {d\chi \over a^2} {\chi' - \chi \over \chi'}
%{\chi'' - \chi \over \chi''} P_{\delta\delta} (k = \ell/\chi)
%S (\chi' < \chi) S (\chi < \chi'')
%\end{eqnarray}
This is the scaling we wish to exploit: for a fixed foreground population,
$W_f$, as one varies the background redshift distribution $W_b$, the 
lensing power
spectra $P_{\rmg\gamma}$ and 
$P_{\gamma\gamma}$ scale in a definite manner,
namely linearly through the factor $1/\chieff(b)$ but with an offset
given by $F$ or $A$ 
(hence the name {\it offset-linear scaling}).  This should
be contrasted with the linear scaling described below.  Moreover, this
factor $1/\chieff(b)$ is purely geometrical. It is the inverse source
distance averaged over the background redshift distribution
(eq.~[\ref{chieff}]).  It is important to emphasize that
eq.~[\ref{approxscaling}] holds even if $W_f$ and $W_b$ are broad distributions
-- the only requirement is that they have little overlap. We will discuss what
requirement this places on the photometric redshift accuracy in
\S\ref{fisher}.  

Such a scaling is very useful in confirming the lensing hypothesis of the
observed correlation \ie intrinsic alignment is not expected to produce this
kind of scaling. This fact can be exploited to weed out contamination of the
observed signals from intrinsic alignment, which will be further explored in a
future paper.

A more ambitious goal is to use this scaling to effectively measure the angular
diameter distance as a function of redshift (more precisely, measure
$\chieff(b)$ as a function of distribution $W_b$), and use this to
constrain cosmological parameters, especially those pertaining to dark energy,
in a way independent of assumptions about the large scale structure of galaxy
and mass. This is the topic of \S\ref{fisher}.

\subsection{Comparison with Linear Scaling}
\label{complinear}

At this point, it is useful to compare the scaling displayed in
eq.~[\ref{approxscaling}] with the scaling used by \cite{jt03}.  Unlike 
offset-linear scaling, the \cite{jt03} scaling can only be applied to 
$P_{\rmg\gamma}$,
\ie to galaxy-galaxy lensing. \cite{jt03} assumed $W_f$ is well
approximated by a delta function at a distance, $\hat{\chi}_\rmf$, in which
case  $G(\ell;f)=-\hat{\chi}_\rmf\,F(\ell;f)$ and $P_{\rmg\gamma}$ follows a
scaling that is even simpler than in eq.~[\ref{approxscaling}] (although
eq.~[\ref{approxscaling}] still holds) \ie {\it a linear scaling} with no
offset:
\begin{eqnarray}
P_{\rmg\gamma} (\ell; f,b)\approx
F(\ell;f)\,\left(1-\frac{\hat{\chi}_\rmf}{\chieff(b)}\right)
\label{JTscaling}
\end{eqnarray}
%\begin{eqnarray}
% P_{\rmg\kappa}(\ell;f,b)\approx
%{3\Omega_{\rmm0} H_0^2 \over 2 c^2}\,{1\over a_\rmf\,\hat{\chi}_\rmf}
%P_{g\delta} (\frac{\ell}{\hat{\chi}_\rmf}) \\ \nonumber 
%\times\int d\chi_\rmb
%W_b(\chi_\rmb)\,{{\chi_\rmb-\hat{\chi}_\rmf}\over\chi_\rmb}
%\end{eqnarray}
Note that all of the uncertainties associated with
large scale structure come in the
prefactor $F(\ell;f)$.  Here the background distribution, $W_b$, does not have
to be well approximated by a $\delta$-function, only the foreground
distribution, $W_f$, does.  One also requires that $W_b$ not extend
significantly into the foreground just as with the offset-linear scaling. For a
fixed foreground redshift, varying the background distribution produces a
definite {\it linear scaling} (with no offset) of $P_{\rmg\gamma}$ with the
geometrical factor $1-\hat{\chi}_\rmf/\chieff(b)$.  \cite{jt03} proposed
that one can examine the ratio of $P_{\rmg\gamma}$'s measured using two
different background distributions ($W_b$ and $W_{b'}$) but the same
foreground: \footnote{We have paraphrased \cite{jt03} a little bit here. They
considered the ratio of correlations measured in real space instead of Fourier
space.} 
\begin{eqnarray}
\label{Pratio}
{P_{\rmg\gamma} (\ell;f,b) \over 
 P_{\rmg\gamma} (\ell;f,b')}
\approx{\hat{\chi}_\rmf^{-1}-\chieff(b )^{-1}
   \over\hat{\chi}_\rmf^{-1}-\chieff(b')^{-1}}.
\end{eqnarray}
One can infer values for cosmological parameters with this equation by
measuring the left-hand-side and then finding the parameters for which the 
right-hand-side yield the same values.

\subsection{The Foreground Width Systematic}
\label{deltafunc}

In practice the foreground galaxies will not have zero uncertainty in
distance, and unless one has spectroscopic redshifts for the foreground
galaxies (\eg McKay et al. 2001, Sheldon et al. 2003), 
$W_f$ will have some non-negligible
spread.  Such a spread implies the ratio of the observed $P_{\rmg\gamma}$'s
will differ from the idealized limit of eq.~[\ref{Pratio}] which can lead to
systematic errors in estimates of cosmological parameters, \eg the dark
energy equation of state, $w$, if one uses the linear scaling but not if one
uses the offset-linear scaling.

If the foreground distribution $W_f$ is not a delta function, 
eq.s [\ref{JTscaling},\ref{Pratio}] should be replaced by:
\begin{eqnarray}
P_{\rmg\gamma} (\ell; f,b)\approx
F(\ell;f)\,\left(1-\frac{\tilde{\chi}_\rmf}{\chieff(b)}\right)
\label{JTscalingOffset}
\end{eqnarray}
\begin{eqnarray}
\label{PratioOffset}
{P_{\rmg\gamma} (\ell;f,b) \over 
 P_{\rmg\gamma} (\ell;f,b')}
\approx{\tilde{\chi}_f(\ell)^{-1}-\chieff(b )^{-1}
   \over\tilde{\chi}_f(\ell)^{-1}-\chieff(b')^{-1}}
\end{eqnarray}
where $\tilde{\chi}_f(\ell)\equiv-G(\ell;f)/F(\ell;f)$.  While
eq.s~[\ref{Pratio},\ref{PratioOffset}] are similar in form the right-hand-side
of the latter is $\ell$ dependent and depends on non-measured and
non-geometrical quantities like the galaxy-mass power spectrum $P_{\rmg\delta}(k,\chi)$.

To bring the non-geometrical character of eq.~[\ref{PratioOffset}] into better
focus, let us perform an expansion of $\tilde{\chi}_f$ around the
mean distance $\bar{\chi}_f \equiv \int d\chi_\rmf W_f(\chi_\rmf)\,\chi_\rmf$
{\it i.e.}
$\tilde{\chi}_f=\bar{\chi}_f+\Delta\chi_{f(2)}+\Delta\chi_{f(3)}+\ldots$
where $\Delta\chi_{f(n)}$ is order $n$ in the width of $W_f$ (the 1st order
term is zero).  The lowest order correction is
\begin{eqnarray}
\label{tildechi}
&& \Delta\chi_{f(2)}(\ell)=-\frac{\sigma_\chi^2}{\bar{\chi}_f} \times \, \\ \nonumber 
&& \left(1-{a_f\,\bar{\chi}_f\,H_f\over c}+n_f(\ell) 
+{2\,a_f H_f\bar{\chi}_f \,\Upsilon_f(\ell)\over c}  \right)
\end{eqnarray}
where $\sigma_\chi$ is the width of $W_f$ {\it i.e.}
$\sigma_\chi^2 \equiv \int d\chi_\rmf W_f(\chi_\rmf) (\chi_\rmf
- \bar{\chi}_f)^2$, 
$a_f\equiv a(\bar{\chi}_f)$, $H_f\equiv H(\bar{\chi}_f)$ is the Hubble
constant at the time corresponding to $\bar{\chi}_f$,
\begin{equation}
\label{SpectralIndex}
n_f(\ell)\equiv
\left.\frac{d{\rm ln}P_{\rmg\delta}(k,\bar{\chi}_f)}{d{\rm ln}k}
                        \right|_{k=\frac{\ell}{\bar{\chi}_f}}
\end{equation}
is the spectral index evaluated at that foreground redshift, and
\begin{equation}
\label{GrowthExponent}
\Upsilon_f(\ell)\equiv {1\over 2}\left.
{d{\rm ln}P_{\rmg\delta}(\frac{\ell}{\bar{\chi}_f},\chi)\over d{\rm ln}a(\chi)}
                        \right|_{\chi=\bar{\chi}_f}
\end{equation}
tells us about the growth of correlations with time.

The terms $n_f (\ell)$ and $\Upsilon_f (\ell)$ 
in $\Delta\chi_{f(2)}(\ell)$ clearly depend on a
non-geometrical quantity, namely the 3-d galaxy-mass
power spectrum $P_{g\delta} (k,\bar\chi_f)$. 
One can imagine improving upon the JT03 procedure by
accounting for corrections due to such terms 
when fitting the ratio of lensing correlations
for dark energy parameters (eq. [\ref{PratioOffset}]).
This somewhat compromises the original goal of
isolating the purely geometrical information.
A more serious problem is that
a quantity like $\Upsilon_f (\ell)$, which
is the growth rate of the galaxy-mass correlation,
is fundamentally rather uncertain because of 
the uncertain relation between galaxy and mass.
Conservatively, this leads to an order $\sigma_\chi^2/\bar\chi_f$ 
uncertainty in any estimate of the correction $\Delta\chi_{f(2)} (\ell)$.

In other words, as long as the foreground distribution $W_f$ has a finite
width, the ratio of correlations considered by JT03 does not give
eq. (\ref{Pratio}), but instead gives eq. (\ref{PratioOffset}), 
where $\tilde{\chi}_f = \bar\chi_f + O(\sigma_\chi^2/\chi_f)$ and
the correction $O(\sigma_\chi^2/\chi_f)$ is uncertain.
Attempts to make use of the JT03 linear scaling to infer
dark energy constraints is therefore subject to a systematic
error that depends on the width of $W_f$. 
Following JT03, consider using a foreground distribution 
by taking a photometric redshift bin centered at for instance $z_p = 0.3$, with
a top-hat width of $\Delta z_p = 0.1$. To obtain the actual 
distribution $W_f$ of {\it true} redshifts, one has to convolve such a 
top-hat photometric redshift bin
with the photometric redshift error distribution, which
we model as a Gaussian of dispersion $\sigma_z$ (this is described
more fully in \S \ref{fisher}). 
We find that the JT03 method (eq. [\ref{PratioOffset}])
is susceptible
to a systematic error of $\sim 30 \%$, $5\%$ or $1\%$
on the dark energy equation of state $w$,
for $\sigma_z = 0.05$, $0.02$ or $0.01$ respectively.
The JT03 linear scaling is therefore quite demanding on
the photometric redshift accuracy if one would like to
keep the systematic error below say $1\%$.
Unless spectroscopic redshifts are available, we think it is
more productive to make use of the offset-linear scaling which
makes no assumptions about the width of $W_f$ and can be
applied to both pure lensing data and galaxy-galaxy lensing.

\subsection{The Ratio of Power Spectrum Differences}

For a zero width foreground galaxy distribution linear scaling means the ratio
of the power spectrum leads to a purely geometric expression
(eq.~[\ref{Pratio}]), while more generally with offset-linear scaling it is the
ratio of difference of power spectra which is purely geometrical:
\begin{eqnarray}
\label{Pdiffratio}
{P(\ell;f,b)  -P(\ell; f,b') \over  P(\ell;f,b'')-P(\ell; f,b''')}
=   {\chieff(b) ^{-1} -\chieff(b')^{-1}
\over\chieff(b'')^{-1}-\chieff(b''')^{-1}}
\end{eqnarray}
where here $P$ can be either $P_{\gamma\gamma}$ or $P_{\rmg\gamma}$.  Here we
illustrate the general case of 4 background populations $b,b',b'',b'''$; but
the expression still gives a non-trivial result for 3 populations, say if
$b=b''$.  Unlike eq.~[\ref{Pratio}] this expression makes no assumptions about
the width of $W_f$ being small.  It does not depend on the mass power spectrum
or its growth, but depends only on the background redshift distributions and
cosmological parameters of interest, such as the equation of state and
abundance of dark energy.

\subsection{The Redshift Tail Systematic}
\label{ztailsystematic}

Another systematic effect which is common to {\it both} linear and
offset-linear scaling comes from the approximations of
eq.s~[\ref{GalGalApprox},\ref{ShearApprox}] that the foreground populations are
completely in front of the background populations.  If this is not true then
the eq.~[\ref{approxscaling}] is not exact, but the exact expression is
\begin{eqnarray}
\label{fullscaling}
P_{\rmg\gamma}  (\ell; f,b)=F(\ell;f)+G(\ell;f)/\chieff(b)+I(\ell;f,b)
                                                                   \\ \nonumber
P_{\gamma\gamma}(\ell; f,b)=A(\ell;f)+B(\ell;f)/\chieff(b)+D(\ell;f,b)
\end{eqnarray}
where the additional terms are given by
\begin{eqnarray}
&& I(\ell; f,b) \equiv{3\Omega_{\rmm0} H_0^2 \over 2 c^2}
\int_0^\infty {d\chi_\rmf\over a(\chi_\rmf)} W_f (\chi_\rmf)  \\ \nonumber
&& \hskip50pt \times\int_0^{\chi_\rmf} d \chi_\rmb W_b (\chi_\rmb)\,
{\chi_\rmf-\chi_\rmb\over \chi_\rmb \chi_\rmf}\,
P_{g\delta}(\frac{\ell}{\chi_\rmf}, \chi_\rmf) \\ \nonumber
&& D(\ell;f,b)\equiv\left({3\Omega_{\rmm0} H_0^2 \over 2 c^2}\right)^2
\int_0^\infty d\chi_\rmf\,W_f (\chi_\rmf) \\ \nonumber 
&&\hskip-15pt \times\int_0^{\chi_\rmf} d\chi_\rmb\,W_b (\chi_\rmb)
 \int_{\chi_\rmb}^{\chi_\rmf}{d\chi \over a(\chi)^2}\,
{\chi-\chi_\rmb\over \chi_\rmb}
{\chi_\rmf-\chi\over\chi_\rmf}\, P_{\delta\delta} (\frac{\ell}{\chi}, \chi)\ ,
\end{eqnarray}
which are both positive (at least so long as $P_{\rmg\delta}>0$).  The ratio of
power spectrum differences is given by eq.~[\ref{Pdiffratio}] only to the
extent that the additional terms $I$ or $D$ are negligible.  
Note that $I$ and $D$ are non-zero only when the foreground distribution
$W_f (\chi_f)$ and background distribution $W_b (\chi_b)$ have 
non-vanishing overlap {\it i.e.}
some of the galaxies identified as foreground are actually behind the
galaxies identified as background ($\chi_f > \chi_b$).
This systematic
effect differs from the foreground width systematic discussed earlier in that it
depends on the tail of the distributions of redshift uncertainties. This is
different from a systematic caused by the width of the foreground distribution
because one can reduce the overlap, and hence the systematic, by selecting the
foreground and background populations in a way which further separates them in
redshift.  Since the tail of the redshift distribution is likely to fall off
rapidly, increasing the separation can greatly decrease the amount of overlap,
and hence the size of $I$ and $D$, and therefore inaccuracy of
eq.~[\ref{Pdiffratio}].  In contrast the foreground width systematic, which
only effect the linear scaling, is not decreased by further separating the
foreground and background populations.  We quantify how large a systematic
error this effect will have on our analysis in \S\ref{fisher}.

\section{A Fisher Matrix Analysis Exploiting the Source Scaling}
\label{fisher}

Here, we would like to find out the dark energy constraints one can in
principle obtain from the offset-linear scaling described in
eq.~[\ref{approxscaling}], using purely geometric quantities like
eq.~[\ref{Pdiffratio}].  Given several redshift bins, one can imagine there are
many ways, or at least many combinations like eq.~[\ref{Pdiffratio}], to obtain
dark energy constraints.  Given a set of $P_{\rmg\gamma}(\ell;f,b)$'s and
$P_{\gamma\gamma}(\ell;f,b)$ 's for a whole range of $f,b$, the best way is
probably to fit them using the offset-linear scaling of
eq.~[\ref{approxscaling}], and marginalize over $A$,$B$,$F$ and $G$.

To estimate the statistical errors, we will assume the mass and
galaxy density fields are approximated by Gaussian
random noise.
On large scales, the near Gaussianity of cosmological
inhomogeneities is quite well established.  Even on small scales where the 3-d
mass and galaxy distribution are far from Gaussian, the projected galaxy and
mass surface density (which gives the shear) are much more
Gaussian since they are a projection of many 3-d structures (\cite{szh},
\cite{white}, \cite{cooray}).  The expected non-Gaussianity will lead to a
small underestimate of errorbars but does not lead to a bias.

To predict the uncertainties in cosmological parameters we use a Fisher matrix
calculation.  For a zero mean Gaussian distribution the Fisher matrix element
for parameters $p_\alpha$ and $p_\beta$ is given by (\eg \cite{max})
\begin{eqnarray}
F_{\alpha\beta}={1\over2}\,
{\rm Tr}\left[\bfC^{-1}\cdot{\partial\bfC\over\partial p_\alpha}\cdot
              \bfC^{-1}\cdot{\partial\bfC\over\partial p_\beta }\right]
\end{eqnarray}
where $\bfC$ is the correlation matrix of the data vector $\bfd$, \ie
$\bfC\equiv\langle \bfd^{\rm T}\bfd\rangle$.  
\footnote{\cite{bj03} also performed a Fisher matrix analysis. 
Their starting point was different from ours, however (in addition to other
differences discussed in Appendix A).  \cite{bj03} started from a likelihood
that treated the power spectra themselves as data which are Gaussian
distributed.} 
Here the elements of $\bfd$ might
consist of local (in angle) shear and galaxy surface density estimators,
however it is more convenient to make linear combinations which, for each
distance bin (foreground or background), are mode amplitudes for approximate
eigenmodes of the angular Laplace operator, with approximate eigenvalue
$-\ell\,(\ell+1)$ (for shear we want only the scalar (E-) eigenmodes).
Discrete sampling by galaxies and incomplete sky coverage will prevent one from
constructing exact eigenmodes in practice, but to estimate the errors it is a
good approximation\footnote{This is true when most of 
the Fisher information comes
from angular scales much smaller than the survey size and much greater than the
typical inter-galaxy separation.} to assume that such modes exist, that the
different angular modes are uncorrelated, and the number of modes with
wavenumber $\ell$ is $(2\ell+1)\,f_{\rm sky}$ where $f_{\rm sky}$ is the
fraction of the sky one has surveyed.  Since the modes are uncorrelated and the
modes for the same angular wavenumber, $\ell$, should have the same
correlations, the correlation matrix $\bfC$ is block diagonal, and we may
rewrite the Fisher matrix element as 
\begin{eqnarray}
F_{\alpha\beta}=f_{\rm sky}\,\sum_\ell (2\ell+1)\,F_{\ell,\alpha\beta}
\end{eqnarray}
where
\begin{eqnarray}
F_{\ell,\alpha\beta}={1\over2}
{\rm Tr}\left[\bfC^{-1}_\ell\cdot{\partial\bfC_\ell\over\partial p_\alpha}\cdot
              \bfC^{-1}_\ell\cdot{\partial\bfC_\ell\over\partial p_\beta }
              \right]\ .
\end{eqnarray}
If there are $n_{\rm bin}$ redshift bins each $\bfC_\ell$ block can be divided 
into $n_{\rm bin}\times n_{\rm bin}$ sub-blocks as follows
\begin{equation}
\label{BIGC}
\bfC_\ell\equiv\left(
\begin{array}{ccc}
\bfC_{\ell,1,1} & \cdots & \bfC_{\ell,1,n_{\rm bin}} \\
\vdots          & \ddots & \vdots                    \\
\bfC_{\ell,n_{\rm bin},1}& \cdots & \bfC_{\ell,n_{\rm bin},n_{\rm bin}}
\end{array} 
\right)\ ,
\end{equation}
one for each ordered pair of distance bins, $(i,j)$.  The sub-blocks are
$2\times2$ matrices given by
\begin{equation}
\bfC_{\ell,i,j}\equiv\left(\begin{array}{cc}
                         P_{\rm gg}(\ell;i,j)+\delta_{ij}{1\over\bar n^g_i} &
                         P_{\rmg\gamma}(\ell; i,j)  \\
                         P_{\rmg\gamma}(\ell; j,i) & 
  P_{\gamma\gamma}(\ell;i,j)+\delta_{ij}{\sigma_{\gamma,i}^2\over\bar n^\rmg_i}
\end{array} \right)
\end{equation}
where $P_{gg} (\ell;i,j)$ is the cross power spectrum at wavenumber
$\ell$ of galaxies in redshift bin $i$ and bin $j$, 
$\bar n^\rmg_i$ is the surface density
of galaxies in redshift bin $i$ which tells us the {\it shot noise}, and
$\sigma_{\gamma,i}^2$ is the intrinsic noise of the shear 
from one galaxy in redshift bin $i$ which tells us the {\it shape noise}.  Since
$P_{\rm gg}(\ell;i,j)=P_{\rm gg}(\ell;j,i)$ and 
$P_{\gamma\gamma}(\ell;i,j)=P_{\gamma\gamma}(\ell;i,j)$ we see that
$\bfC_{\ell,i,j}=\bfC_{\ell,j,i}^{\rm T}$ and that $\bfC_\ell$ is symmetric.
If the redshift bins are reasonably large, it will be a good
approximation to ignore galaxy correlations between bins, \ie we assume
$P_{\rm gg}(\ell;i,j)=\delta_{ij}P_{\rm gg}(\ell;i)$.

We suppose that the galaxies in each population have measured {\it photometric}
redshifts, $z_{\rm p}$ which estimate the true redshift.  We assume that
the distribution of $z_{\rm p}$ from all galaxies in all bins is 
\begin{eqnarray}
\label{dNdz}
{d N_\rmp(z_\rmp)\over dz_\rmp}\propto 
   z_\rmp^2\,e^{-\left({z_\rmp\over 1.0}\right)^{1.5}}\ .
\end{eqnarray}
We divide this total population into $n_{\rm bin}$ top-hat bins in
$z_\rmp$-space, such that bin $i$ contains all galaxies with 
$(i-1)\,\Delta z_\rmp\le z_\rmp<i\,\Delta z_\rmp$. We suppose the
photometric redshifts are unbiased estimators of the true redshift with errors
distributed like a Gaussian with variance $\sigma_z^2$ so that the distribution
of true redshifts in bin $i$ is
\begin{eqnarray}
\label{Witrue}
{d N_i(z)\over dz}\propto
\int_{(i-1)\,\Delta z_{\rm p}}^{i\,\Delta z_{\rm p}}dz_\rmp\,
{d N_\rmp(z_\rmp)\over dz_\rmp}\,
e^{-{1\over 2}\left({z-z_\rmp\over\sigma_z}\right)^2}
\end{eqnarray}
which then is related to $W_i(\chi)$ through eq.~[\ref{DistanceDistribution}]
and from that we can compute, for a given set of cosmological parameters, the
effective distance to each bin, $\chieff(i)$, from eq~[\ref{chieff}].

Note that we have not yet defined foreground and background bins or exploited
offset-linear scaling.  To do so we define foreground/background pairs by the
requirement that bin $j$ is a background bin to bin $i$ if $j\ge i+\Delta_{\rm
bin}$.  If $b=j$ is a background bin to foreground bin $f=i$ then
$P_{\rmg\gamma}(\ell;f,b)$ and $P_{\gamma\gamma}(\ell;f,b)$ are given by
eq.~\ref{approxscaling} while $P_{\rmg\gamma}(\ell;b,f)=0$ and
$P_{\gamma\gamma}(\ell;b,f)=P_{\gamma\gamma}(\ell;f,b)$.  One minimally
requires that $\Delta_{\rm bin}=1$, however in this case one is subject to 
systematics by the redshift tail to the extent that the redshift distributions
between adjacent bins overlap. Increasing
$\Delta_{\rm bin}$ decreases any systematic effect from redshift tails, however
it also leads to larger statistical errors because more information is thrown away.
We will discuss below the choice of $\Delta_{\rm bin}$ and redshift binning.

The correlation matrix depends on several functions having to do with the power
spectrum: there are the foreground  functions $A(\ell;f)$, $B(\ell;f)$,
$F(\ell,f)$, and $G(\ell,f)$; and then there are the lensing power spectra 
$P_{\rmg\gamma}(\ell;i,j)$ and $P_{\gamma\gamma}(\ell;i,j)$ where neither
$\{i,j\}$ nor $\{j,i\}$ forms a foreground-background pair (as defined above
via $\Delta_{\rm bin}$); finally there are the galaxy angular
power spectra $P_{gg} (\ell; i,j)$. These are
functions we are not interested in because we are only interested in obtaining
constraints on dark energy properties which are independent of the values of
these functions. So we assume their values are not known {\it a priori}, and
for each $\ell$ we take their values to be unknown {\it nuisance parameter}s,
\ie each corresponding to one component of $p_\alpha$. So the number of unknown
parameters will be very large, but since dependence on each of these
uninteresting parameters is confined to a single block, $\bfC_\ell$, the
computation of the Fisher matrix remains tractable.

In addition to the nuisance parameters the correlation matrix depends on
the background bin distances: $\chieff(b)$ (eq. [\ref{chieff}]). These will depend on interesting
cosmological parameters through the function $\chi(z)$ and
eq.~[\ref{DistanceDistribution}].  The cosmological parameters we are actually
interested in are: $w$ the equation of state of dark energy, $w' \equiv dw/dz$, 
and $\Omega_{\rm de}$ the dark energy density today in unit of the
critical density.  We assume a flat universe here, so the matter density is
given by $\Omega_{\rmm0}=1-\Omega_{\rm de}$.

To remove the nuisance parameters we can marginalize over their values.  This
can be done by inverting the full Fisher matrix, $F_{\alpha\beta}$, and then
restricting the inverse to the interesting cosmological parameters, let us
denote them by $\tilde\alpha, \tilde\beta$:\footnote{In
order for $F_{\alpha\beta}$ to be invertible we require that
$n_{\rm bin}\ge\Delta_{\rm bin}+3$ otherwise there are never the 3 background
bins required to construct the ratio of power spectrum differences,
eq.~[\ref{Pdiffratio}], so that one can make use of the offset-linear scaling.}
\begin{eqnarray}
\widetilde E_{\tilde\alpha \tilde\beta} \equiv (F^{-1})_{\tilde \alpha \tilde \beta}
\end{eqnarray}

According to the Cramer-Rao inequality the minimum possible error ellipses in
parameter space (for unbiased estimators) have principal axes in the directions
of the eigenvectors of $\widetilde{E}_{\tilde\alpha\tilde\beta}$ with size given by the square
root of the 
corresponding eigenvalues of $\widetilde{E}_{\tilde\alpha\tilde\beta}$.
Maximum likelihood parameter estimators (MLEs) will approach this accuracy
where the errorbars are small enough. For the problem at hand we expect these
minimum errors to close to what can be obtained in practice.

%\footnote{There are actually 2 extra complications.
%The first is that $P_{\rmg\kappa} (\ell; i,j)$ and 
%$P_{\kappa\kappa}(\ell;i,j)$
%should be treated as free parameters to be marginalized over, when $i=j$. This
%is because the offset-linear scaling of eq.~[\ref{approxscaling}] does not
%hold when $i=j$, due to a non-negligible $D(\ell; i,j)$ and $H(\ell;i,j)$.
%The
%second is that, suppose the total number of redshift bins is $N_z$, then
%$P_{\rmg\kappa} (\ell; N_z-1, N_z)$ and $P_{\kappa\kappa}(\ell;N_z-1,N_z)$
%should be treated as free parameters to be marginalized over as well (along
%with this, $A(\ell; N_z-1)$, $B(\ell;N_z-1)$, $F(\ell;N_z-1)$ and
%$G(\ell;N_z-1)$ can be removed from the list of free parameters). Why this
%should be so is easiest to see if $N_z = 2$.  In that case, there is simply
%not enough information to exploit the offset-linear scaling, and fit for all
%relevant parameters.}

To obtain prediction for how accurately one can constrain cosmological
parameters using offset-linear scaling, we take $f_{\rm sky}=0.1$, (\ie a
$4000\,(^\circ)^2$ survey), $\sigma_{\gamma,i}^2=0.3^2/2$ (shape noise), and
$\sum_{i=1}^{n_{\rm bin}}\bar{n}^\rmg_i=100/(')^2$. 
For the fiducial cosmological and structure formation model, we use
$w = -1$, $\Omega_{\rm de} = 0.7$, a scale invariant primordial mass power 
spectrum with a linear amplitude of $\sigma_8 = 0.9$, 
and for the galaxy and galaxy-mass power spectra, we employ
the halo model (Sheth \& Jain 1997, 
Ma \& Fry 2000, Seljak 2000, Scoccimarro et al. , Guzik and Seljak 2001).
To distribute galaxies inside halos, we use the occupation function given
by Kratsov et al. (2003), with a galaxy (subhalo) masscut at each redshift that matches
the redshift distribution given in eq. [\ref{dNdz}] with
a total integrated number density of $\sum_{i=1}^{n_{\rm bin}}\bar{n}^\rmg_i=100/(')^2$.
We have experimented with using only the more massive halos as foreground following JT03, but
found it did not lead to an improvement in statistical errors.

\begin{figure}[tb]
\centerline{\epsfxsize=9cm\epsffile{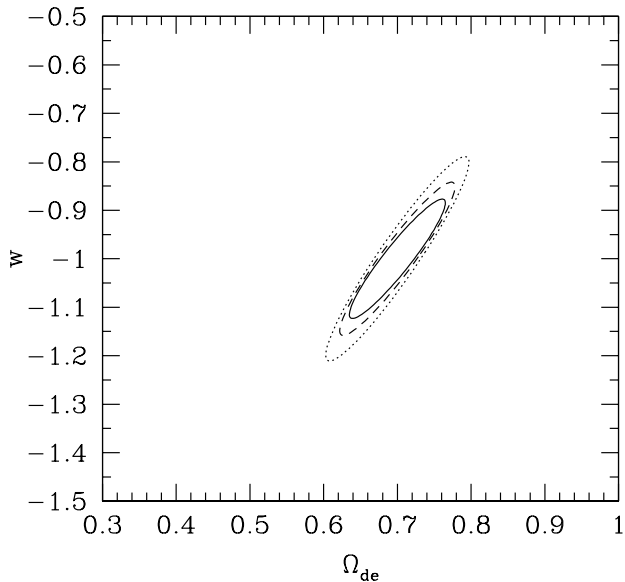}}
\caption{Purely geometrical weak-lensing
constraints on $w$ and $\Omega_{\rm de}$, making use of the
offset-linear scaling of eq.~[\ref{approxscaling}], for
a survey covering $10 \%$ of the sky, and median redshift $\sim 1$
(eq. [\ref{dNdz}]).
The solid, dashed and dotted (1 $\sigma$) contours correspond to
a photometric redshift error of $\sigma_z = 0.01, 0.02$ and $0.05$
respectively. The size of $\sigma_z$ is relevant only because
the redshift bins  are chosen to keep systematic error below $1\%$, 
and the statistical error is affected by the redshift binning (see text).
The equation of state $w$ is assumed
constant. No prior is placed on $\Omega_{\rm de}$
or $w$. The fiducial model has $w = -1$ and $\Omega_{\rm de} = 0.7$.
}
\label{OmegaVw_20}
\end{figure}

\begin{figure}[tb]
\centerline{\epsfxsize=9cm\epsffile{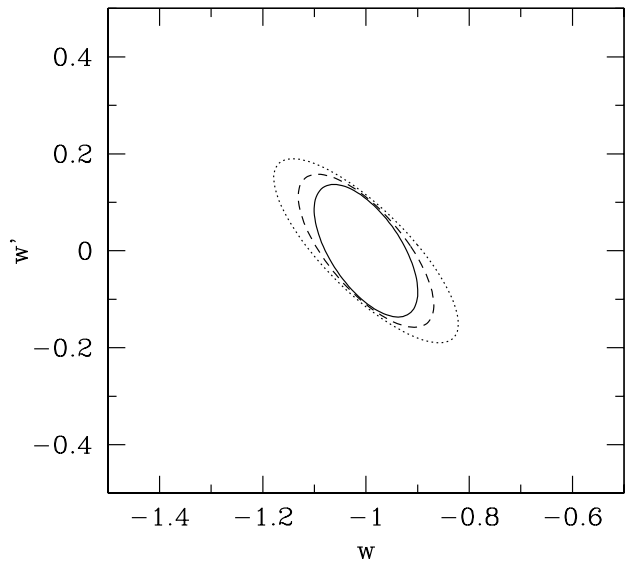}}
\caption{Purely geometrical weak-lensing
constraints on $w$ and $w'$, making use of the
offset-linear scaling of eq.~[\ref{approxscaling}], for
a survey covering $10 \%$ of the sky, and median redshift $\sim 1$
(eq. [\ref{dNdz}]).
The solid, dashed and dotted contours have the same meanings
as in Fig. \ref{OmegaVw_20}. A $3 \%$ uncertainty on 
$\Omega_{\rm de}$ is assumed.
The fiducial model has $w = -1$, $w' = 0$ and $\Omega_{\rm de} = 0.7$, where
$w' \equiv dw/dz$. Note that a common alternative
parametrization has $w_a = 2 w'$ at $z=1$ (see text).
}
\label{Omegaww_prime_20}
\end{figure}

Carrying out the Fisher matrix calculation as outlined above we obtain
the dark energy constraints as shown in Fig.~\ref{OmegaVw_20} and
Fig.~\ref{Omegaww_prime_20} which show, respectively the constraints on
$\Omega_{\rm de}$ and $w$ when $w'=0$,
and the constraints on $w$ and $w'$ when
$\Omega_{\rm de}$ is assumed to be known to $3\%$.  
The symbol $w'$ denotes $dw/dz$. A common alternative parametrization
of evolution of $w$ has $w_a = 2w'$ at $z=1$ (Linder 2002).
The solid, dashed and dotted contours
give 1 $\sigma$ errors that correspond to a photometric redshift accuracy
of $\sigma_z = 0.01, 0.02$ and $0.05$ respectively.
For each photometric redshift accuracy, we choose a redshift binning
that keeps the redshift tail systematic error
(\S \ref{ztailsystematic}) at a sub-percent level
while minimizing the statistical error.
(Keeping the systematic error on dark energy parameters
at a sub-percent level is
probably more stringent than is necessary given the size of
the statistical error as it turns out.)
For a photometric redshift error of $\sigma_z = 0.05$, we choose 
$(\Delta z_\rmp, n_{\rm bin}, \Delta_{\rm bin}) = 
(0.15, 20, 2)$; 
for $\sigma_z = 0.02$, we choose $(0.1, 30, 2)$, and for
$\sigma_z = 0.01$, we consider $(0.15, 20, 1)$. 
The values of $\sigma_z$ considered should span a reasonable range
of what can be achieved with photometric redshifts. The redshift bins stretch
out to $z\sim3$ which encompasses the redshift range of most normal galaxies.
Sampling photometric redshift space more finely by decreasing $\Delta z_\rmp$
while increasing $n_{\rm bin}$ accordingly gives negligible improvement in
errorbars mainly because of the increased importance of shot-noise when the bin
size is made small (see \cite{hu99}).

%From /merton1/lhui/Newfermi/Corr/Jun/Fig12_10_03b/z1.0/OmegaVw.ps

%old version, with 1 % prior on Omega_de: /merton1/lhui/Newfermi/Corr/Jun/Fig12_10_03b/z1.0/Omegaww.ps
%the current version, with 3% prior: 
%/merton1/lhui/Newfermi/Corr/Jun/Fig12_11_03/z1.0/sigOmega0.03/Omegaww.ps
%		or at laptop: GeomSep5/Fig2butwith0.03/z1.0/sigOmega0.03/Omegaww.ps

Note that because we marginalize over all parameters that are determined by the
mass, galaxy and galaxy-mass power spectra, it is sensible for us to use
information from all scales (from the fundamental mode to 
$\ell \sim 10^5$)-- \ie there is no need to stay away from nonlinear
scales because of worries about how well one can predict the mass and galaxy
power spectra.  Of course, at sufficiently high $\ell$'s, shape-noise
dominates and not much information is gained from the very high $\ell$ modes.

Fig.s~\ref{OmegaVw_20} \& \ref{Omegaww_prime_20} show that our geometrical
method employing offset-linear scaling yields weaker dark energy constraints than
conventional weak lensing methods which make more
assumptions about the structure formation model ($w$ generally constrained
at the few percent level for a comparable survey as above; 
see {\it e.g.} Hu 2002, Abazajian \& Dodelson 2003).
This is of course not surprising, since the offset-linear scaling method
throws away non-geometrical information that is utilized in conventional methods.
However, the geometrical constraints are still sufficiently tight to
provide an interesting consistency check: dark energy constraints obtained
using the two different methods should agree;
disagreement would point to flaws in the structure formation model assumed,
or to systematic errors in the data.

Note that our constraints are a bit weaker than 
those obtained by \cite{jt03} and \cite{bj03}.
One might think this could
be due to the fact that we use the offset-linear scaling rather than the 
Jain-Taylor linear scaling -- the former involves more parameters
than the latter (compare eq. [\ref{approxscaling}] and [\ref{JTscaling}]). 
On the other hand, the offset-linear scaling allows the use of both
shear-shear and galaxy-shear correlations while the Jain-Taylor linear scaling
can be applied only to galaxy-shear. So, it is not {\it a priori} obvious
how our constraints should compare with those of JT03 and BJ03.
In Appendix A, we will discuss what happens if we carry
out parameter estimation using the linear scaling. 
We find dark energy constraints that are weaker than those
obtained by \cite{jt03} and \cite{bj03} even in that case.
The reasons are discussed in Appendix A.

\section{Geometry as a Consistency Check}
\label{2ws}

Our procedure, described in the last section, making use of the offset-linear
scaling of eq.~[\ref{approxscaling}], is very conservative \ie we marginalize
over all possible 3-d mass, galaxy, and galaxy-mass power spectra in order
to extract the pure geometrical information.
In truth, we do know a fair amount about these power spectra, especially
from non-lensing observations. The conventional approach is to 
assume the 3-d mass power spectrum is well constrained from other observations
(such as the microwave background),
and fit for dark energy constraints from shear-shear correlations
which depend on dark energy parameters through both geometrical distances
and the growth rate of the mass power spectrum 
(which of course implicitly assumes a structure formation model such
as Cold-Dark-Matter; see {\it e.g.} Hu 2002, Abazajian \& Dodelson 2003).
In other words, unlike the offset-linear scaling method which introduces
a whole set of nuisance parameters in addition to dark energy parameters, 
the conventional approach has {\it only} the dark energy parameters as free parameters. 
A simple alternative, which is less conservative than the offset-linear scaling method,
but allows a consistency test that the conventional approach does not
offer, is as follows.
Follow the conventional approach, but
split the dark energy parameters into
two kinds: those that enters the growth factor, and those that enters the geometrical
distances, and fit for these separately.
With such parameter-splitting (Stebbins 2003), one does not expect and 
will not obtain better constraints compared
to the conventional approach where equivalence between these two sets of parameters
is enforced. The rationale for parameter-splitting is to check for consistency:
if we could verify that the values of $w$ for example obtained separately
from geometry and from growth (let us call them $w({\rm geometry})$ and
$w({\rm growth})$) are consistent with each other, this would increase
our confidence in the values obtained; if they disagree, the discrepancy would
help isolate what was going wrong, say systematic errors (\eg \cite{h03}), 
or contamination of shear by intrinsic alignments
(\cite{lp00,cm00,hrh00,bthd00,ckb01,mwk01,p01a,vv02,maller01,vB02,hz02}), or
incorrect assumptions about the mass power spectrum.

\begin{figure}[tb]
\centerline{\epsfxsize=9cm\epsffile{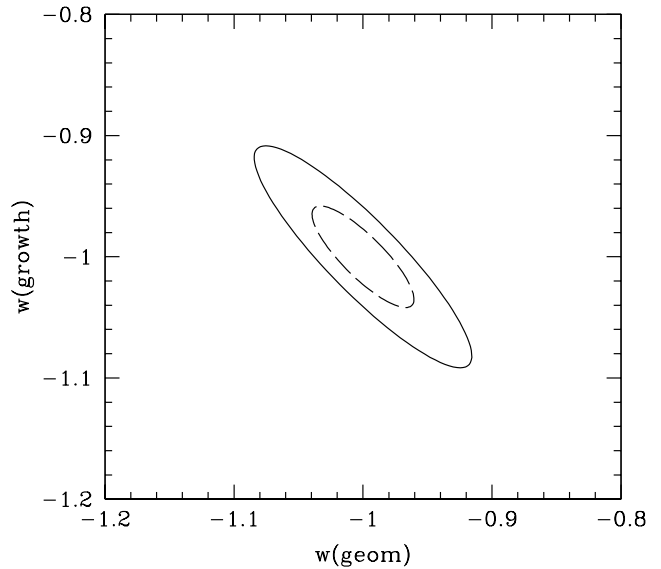}}
\caption{Constraints on the dark energy equation of state $w$
following the conventional approach ({\it i.e.} not
using offset-linear scaling; see text), but with the parameter split
into two: one controlling the growth factor and the other controlling
the geometrical distances. 
The solid (1$\sigma$) contour is from using just the shear-shear power spectrum
$P_{\gamma\gamma}$. The dashed (1 $\sigma$)  contour is from using both shear-shear
and galaxy-shear power spectra $P_{g\gamma}$. $\Omega_{\rm de}$ is fixed at
$0.7$. The survey size and depth are the same as those of 
Fig. \ref{OmegaVw_20}.
}
\label{split}
\end{figure}

As an illustration, in Fig. \ref{split}, we show such a consistency test via
parameter-splitting. We adopt the same fiducial model as in Fig. \ref{OmegaVw_20}, 
and estimate the constraints on $w({\rm geometry})$ and $w({\rm growth})$ from
both the shear-shear power spectrum $P_{\gamma\gamma}$ and the galaxy-shear
power spectrum $P_{g\gamma}$. To fit the galaxy-shear data, we assume
the galaxies are linearly biased with respect to the mass, and we marginalize
over an independent galaxy-bias for each redshift bin ($n_{\rm bin} = 20$).
We limit ourselves to information from $\ell < 200$, for two reasons:
the galaxy-bias is probably not linear on smaller scales; the nonlinear mass
power spectrum might not be accurately predicted even though we assume here the linear
mass power spectrum is well constrained from other observations.
Fig \ref{split} shows that such a consistency test can yield constraints
that are interesting precision-wise.
It is also interesting how using both $P_{\gamma\gamma}$ and $P_{g\gamma}$
gives significantly better constraints than using just one of them.

\section{Discussion}
\label{discuss}

In this paper, we have introduced a special scaling, which we call the
offset-linear scaling: imagine one has a foreground population of galaxies from
which one forms either a galaxy-density field or a shear field;
when one cross-correlates this field with the shear measured from
some background population, the cross-correlation signal ($P_{\rmg\gamma}$ or
$P_{\gamma\gamma}$) scales with the redshift of the background population in a
way that is specific to lensing.  This is the content of
eq.~[\ref{approxscaling}].  Such a scaling can be exploited to extract purely
geometrical information from a lensing survey. Effectively, one can measure
angular diameter distances (or more accurately, combinations thereof;
eq.~[\ref{Pdiffratio}]) from a lensing experiment without making any
assumptions about the shape or growth of the mass/galaxy power spectrum. The idea is
to measure the galaxy-shear and shear-shear power spectra,
$P_{\rmg\gamma}$ and $P_{\gamma\gamma}$ for a variety of foreground and
background redshift bins.  Given a sufficient number of bins, one can fit for
all the quantities $A$, $B$, $F$, $G$ and $\chieff$ in
eq.~[\ref{approxscaling}]. One can then obtain dark energy constraints from
$\chieff$ alone, which is a purely geometrical quantity, essentially an angular
diameter distance weighed in a particular way (eq.~[\ref{chieff}]).

Such an approach has certain virtues. The obvious one is that the resulting
constraints are free of assumptions about one's structure formation model
(typically a Cold Dark Matter model with a nearly scale invariant primordial 
power spectrum). Because of this, one can also make use of information on
smaller scales than what one would otherwise feel uncomfortable using, either
because of non-linearity in the case of $P_{\gamma\gamma}$,
or because of galaxy-biasing in the case of $P_{\rmg\gamma}$. 

The level of constraints from this method employing the offset-linear scaling 
is shown in Fig. \ref{OmegaVw_20} and Fig.~\ref{Omegaww_prime_20}.
The constraints are weaker than conventional methods such as lensing tomography
(Hu 2002, Abazajian \& Dodelson 2003). This is not surprising since the offset-linear
scaling method isolates and uses only the geometrical information, whereas
conventional methods make use of information from both growth and geometry and 
assumes the mass power spectrum is well constrained from other methods.
Nonetheless, the constraints are sufficiently interesting to 
make our geometrical method a useful consistency check on assumptions
behind conventional methods ({\it e.g.} Cold Dark Matter structure formation model).
Comparing against \cite{jt03} and \cite{bj03}, who used a similar geometrical
approach as here but a different scaling (we call the linear scaling; eq. [\ref{JTscaling}]), 
it appears our constraints are weaker than theirs. We believe the reason is
largely because the statistical errors have been underestimated by
\cite{jt03} and \cite{bj03}. This is discussed in detail in Appendix A.

A useful feature of the offset-linear scaling is that it is not as demanding
on the photometric redshift accuracy as the linear scaling (see \S \ref{deltafunc}, 
\ref{ztailsystematic}). 
Another useful feature: the offset-linear scaling can be applied to both shear-shear 
and galaxy-shear correlations, whereas the linear scaling can be applied
only to the latter (see \S \ref{complinear}). 

In \S\ref{2ws}, we introduce the idea of parameter splitting.
In fitting for dark energy parameters to the observed lensing power
spectra (as done in the conventional approach), one 
can artificially split them up into those that control the
growth factor, and those that control the geometrical distances. Consistency
between the two sets would be a good check for the presence of systematic
errors, intrinsic alignment or incorrect assumptions about the nature of the
mass fluctuations. This consistency test is less conservative than
the one using the offset-linear scaling. 

In a sense, the techniques outlined in \S\ref{scaling} and \S\ref{2ws}
represent two extremes of a whole spectrum of ways to separate geometrical
information from growth information: from making no assumptions about the mass
(and galaxy-mass) power spectrum to assuming that it is known to high
precision.  There are likely techniques that are intermediate in this spectrum
that might also prove useful.

We are grateful to Gary Bernstein, Bhuvnesh Jain, and especially Wayne Hu
for useful discussions.
Support for this work is provided in part by the DOE and its Outstanding Junior
Investigator Program, by NASA grant NAG5-10842 and NSF grant AST-0098437. LH is
grateful to the organizers of the superstring cosmology program at the KITP,
where part of this work was done.

\vskip 1cm

\section*{Appendix A -- Comparison with JT03}
\label{appendixA}

Our aim in this Appendix is to discuss 
our differences from Jain \& Taylor (2003) [\cite{jt03}]
and to a lesser extent Bernstein \& Jain (2003) [BJ03].

We all share the common goal of isolating geometrical constraints
on dark energy from lensing data. JT03/BJ03 focused on the
use of the linear scaling (eq. [\ref{JTscaling}]) while we focus
on the offset-linear scaling (eq. [\ref{approxscaling}]). 
The linear scaling introduces fewer nuisance parameters but can only be
applied to galaxy-shear, not shear-shear, correlation data.
It is therefore not {\it a priori} obvious whose constraints should be stronger.
The most direct comparison
can be made between the solid contour of our Fig. \ref{OmegaVw_20}
and the smallest contour in Fig. 1 of JT03\footnote{The method of JT03 requires
high photometric redshift accuracy, hence the $\sigma_z = 0.01$ contour
of our Fig. \ref{OmegaVw_20} is the relevant one to compare against.}.
Our constraints appear to be weaker by about a factor of 3 compared to JT03.
What is puzzling is that even when we adopt exactly the JT03 linear scaling,
and redo our calculation, the constraints are still weaker than those of JT03 by 
a factor of at least 3 or more. (The discrepancy depends on exactly how the JT03 scaling
is implemented, particularly on the choice of redshift bins;
the choice of bins in JT03 seems to lead to statistical errors larger than factor of 3, see below).
This translates into at least an order of magnitude difference in the variance.
This is not a small discrepancy, particularly when we use exactly the JT03 method.
In this Appendix, we will focus on this discrepancy with JT03, but will also
briefly comment on the treatment of BJ03 (who obtained similar constraints as JT03).

We believe the statistical error quoted in JT03 have been underestimated.
There appears to be several different reasons, the first two of which
were pointed out to us by Wayne Hu (see Hu \& Jain 2003).
First, JT03 adopted a singular isothermal spherical profile for cluster halos
that they considered. More realistic profiles such as NFW produce a smaller
lensing signal. Second, it appears profile aside,
the lensing signal itself is overestimated.
Third, which is the aspect we would like to focus on, we believe not all sources
of statistical errors were taken into account by JT03.
Hu \& Jain (2003) also independently reached the same conclusions.

To recapitulate, \cite{jt03} proposed to examine the ratio of
the galaxy-shear
correlation at two different redshifts. For simplicity, we will
consider the ratio of the galaxy-convergence correlation instead,
which can of course be obtained from the galaxy-tangential-shear correlation:
\footnote{In previous sections of the paper, we have been loosely using
the term shear $\gamma$ as equivalent to convergence. In the appendix here, to avoid
confusion especially in Appendix B, we explicitly use the symbol $\kappa$ when
we are discussing convergence.}
\begin{eqnarray}
\label{Ri}
R^{i} = P_{\rmg\kappa}^{i,1} / P_{\rmg\kappa}^{i,2}
\end{eqnarray}
where $i$ specifies some foreground population, and
$1$ and $2$ refers to convergence from 2 different
background redshift bins.
We use the symbol $P_{\rmg\kappa}$ loosely to refer
to either galaxy-convergence correlation function, or
the galaxy-convergence power spectrum. Which is which should
be clear from the context (actual power spectrum will
usually have argument $\ell$).
\footnote{\cite{jt03} actually considered halo-shear rather
than galaxy-shear. We will continue to use the term
galaxy-shear. All our expressions are equally valid for special
classes of foreground 'galaxies' such as groups or clusters.}

The statistical error on dark energy parameters clearly
comes from the statistical error on $R^{i}$, which in turn
is determined by the statistical error of the $P_{\rmg \kappa}$ 
correlations. Before launching onto a detailed calculation, 
it is helpful to indicate roughly where we disagree with JT03
(and also BJ03). 
Think of $P_{\rmg\kappa}$ as $\sim \langle \delta_g \kappa \rangle$.
Its variance under Gaussian random approximation (relaxing the Gaussian
assumption would only increase the error) should
be $\langle \delta_g \kappa \delta_g \kappa \rangle
- \langle \delta_g \kappa \rangle \langle \delta_g \kappa \rangle
\sim \langle \kappa \kappa \rangle \langle \delta_g \delta_g \rangle 
+ \langle \delta_g \kappa \rangle \langle \delta_g \kappa \rangle$.
As we will argue, what JT03 appeared to have considered is only the 
part of the variance that comes from the product of shape-noise in 
$\langle \kappa \kappa \rangle$, and shot-noise in 
$\langle \delta_g \delta_g \rangle$ {\it i.e.} $\sigma_\kappa^2 / (\bar n^B \bar n^g)$,
where $\sigma_\kappa^2$ is the shape-noise of each background galaxy, 
$\bar n^B$ is the number density of background and $\bar n^g$ is the number density
of foreground galaxies. 
In other words, JT03 appeared to have ignored sampling variance terms.
Not only do these terms ignored by JT03 increase the variance of
the measured $P_{\rm g\kappa}$ (and $R^i$), they also introduce correlation
in errors between $R^i$'s measured from different foreground bins, which was
also absent in JT03.

Let us now derive the errorbar on $R^i$ in detail.
The estimator for $P_{\rmg\kappa}$ can be written as
\begin{eqnarray}
\hat P_{\rmg\kappa} = \sum_{\alpha\beta} \delta^g_\alpha \kappa_\beta
\tilde W_{\alpha\beta}
\label{Pgshat}
\end{eqnarray}
The picture in mind is to think of the survey being divided into
pixels, and 
$\delta^g_\alpha$ is the galaxy overdensity in pixel $\alpha$, while
$\kappa_\beta$ is the convergence in pixel $\beta$.
The symbol $\tilde W_{\alpha\beta}$ can stand for
many different things. For example,
if one is interested in the real space correlation function 
at separation $\Delta \theta$, 
$\tilde W_{\alpha\beta}$ should be equal to zero when
the separation between $\alpha$ and $\beta$ differs from $\Delta \theta$, 
or else equal to $1/N$, where $N$ is the total number of pairs of pixels
at that separation. 
If one is interested in the power spectrum at wavenumber $\ell$,
$\tilde W_{\alpha\beta} = (1/ N_{\rm pix}^2) {\,\rm exp} 
(-i \ell \cdot \Delta \theta_{\alpha\beta})$
where $N_{\rm pix}$ is the total number of pixels, and
$A_T$ is the total survey area.
\footnote{Strictly speaking, one is usually interested in 
the power spectrum at a given $|\ell |$, and so $\tilde W_{\alpha\beta}$
should involve an average over directions of $\ell$. 
Note also that the $\tilde W_{\alpha\beta}$ differs from the
conventional one by a factor of $A_T$, but that is fine since
we are only interested in fractional error. Our choice
is to enforce $\sum_{\alpha\beta} \tilde W_{\alpha\beta} = 1$,
which simplifies some of our expressions below.
}
\cite{jt03} considered a particular $\tilde W_{\alpha\beta}$ that
corresponds to averaging the galaxy-convergence correlation
over some aperture. We will keep $\tilde W_{\alpha\beta}$ general
for now.

One word about the estimator $\hat P_{g \kappa}$. 
It might appear very different from the way one usually thinks
of galaxy-galaxy lensing. The usual approach is to 
sit on a foreground galaxy, measure the background tangential
shear averaged around a circle, then average over all foreground galaxies
(Brainerd, Blandford \& Smail 1996, Fischer et al. 2000, McKay et al. 2001).
This is equivalent to
measuring $\sum_{\alpha\beta} (n^g_\alpha/\bar n^g) \gamma^t_\beta 
\tilde W_{\alpha\beta}$
where $\gamma^t$ is the background tangential shear, and
$n^g_\alpha$ is equal to unity if pixel $\alpha$ has a foreground galaxy
or vanishes otherwise, and $\bar n^g$ is its average over the survey.
It is easy to see that such an estimator on average is equivalent
to $\sum_{\alpha\beta} \delta^g_\alpha \gamma^t_\beta 
\tilde W_{\alpha\beta}$, where $\delta^g_\alpha = n^g_\alpha / \bar n^g - 1$. 
The only difference between this and the estimator in
eq.~[\ref{Pgshat}] is the replacement of $\gamma^t$ by $\kappa$. 
This is merely for the sake of simplifying our following expressions.
Finally, note that using $\delta^g$ in place of $n^g/\bar n^g$ is generally
a good idea because it reduces the variance
of the estimator (Szapudi \& Szalay 1998).

The estimator for $R^i$ is given by
\begin{eqnarray}
\hat R^i = \hat P_{\rmg\kappa}^{i,1} / \hat P_{\rmg\kappa}^{i,2}
\label{Rhat}
\end{eqnarray}
We caution here that the above estimator is unbiased only
to the lowest order in fluctuations, but we will ignore
such complications here (\eg \cite{hg99}).

Eq.s~[\ref{Pgshat},\ref{Rhat}] imply the following
expression for the fractional variance of the ratio $R^i$:
\begin{eqnarray}
\label{VR}
V (i) \equiv \langle (\delta \hat R^i)^2 \rangle 
/ (R^i)^2 = V^1 (i) + V^2 (i) - 2 V^{1,2} (i)
\end{eqnarray}
where
$V^1$ is the fractional variance of $\hat P_{\rmg\kappa}^{i,1}$,
$V^2$ is the corresponding quantity for $\hat P_{\rmg\kappa}^{i,2}$,
and $V^{1,2}$ is the cross-variance between them, and
they are given by (approximating fluctuations
as Gaussian random):
\begin{eqnarray}
\label{VRs}
&& V^1 (i) = 
[\int {d^2 \ell \over (2\pi)^2} P_{\rmg\kappa}^{i,1} (\ell) J(\ell)]^{-2} 
\times \\ \nonumber 
&& \int {d^2 \ell \over (2\pi)^2} {|J_\ell |^2 \over A_T}
[ P_{\rmg\kappa}^{i,1} (\ell)^2 + (P_{gg}^{i,i} (\ell) + {1\over \bar n^g_i}) 
(P_{\kappa\kappa}^{1,1} (\ell) + {\sigma_\kappa^2 \over 
\bar n^B_{1}}  )] \\ \nonumber
&& V^2 (i) = 
[\int {d^2 \ell \over (2\pi)^2} P_{\rmg\kappa}^{i,2} (\ell) J(\ell)]^{-2} 
\times \\ \nonumber 
&& \int {d^2 \ell \over (2\pi)^2} {|J_\ell |^2 \over A_T}
[ P_{\rmg\kappa}^{i,2} (\ell)^2 + (P_{gg}^{i,i} (\ell) + {1\over \bar n^g_i}) 
(P_{\kappa\kappa}^{2,2} (\ell) + {\sigma_\kappa^2 \over 
\bar n^B_{2}}  )] \\ \nonumber
&& V^{1,2} (i) = [\int {d^2 \ell \over (2\pi)^2} P_{\rmg\kappa}^{i,1} (\ell) 
J(\ell)]^{-1} \times \\ \nonumber &&
\quad \quad \quad \quad 
[\int {d^2 \ell \over (2\pi)^2} P_{\rmg\kappa}^{i,2} (\ell) J(\ell)]^{-1}
\times 
\\ \nonumber && 
\int {d^2 \ell \over (2\pi)^2} {|J_\ell |^2 \over A_T}
[P_{\rmg\kappa}^{i,1} (\ell)  P_{\rmg\kappa}^{i,2} (\ell) + 
(P_{gg}^{i,i} (\ell) + {1\over \bar n^g_i}) P_{\kappa\kappa}^{1,2} (\ell)]
\end{eqnarray}

Here, $P_{\rmg\kappa}^{i,1}$ is the power spectrum between galaxies
in foreground bin $i$ and convergence in the background bin $1$ (there
are only 2 background bins in JT03), 
$P_{gg}^{i,i}$ is the power spectrum of foreground galaxies with themselves
in bin $i$, and so on. The symbol $\sigma_\kappa^2$ represents
the variance in convergence due to the intrinsic noise of each galaxy,
and $\bar n^g_i$ is the galaxy density in foreground bin $i$, 
$\bar n^B_{1}$ is the density of galaxies in background bin $1$,
and so on. The total survey area is $A_T$.
The quantity $J(\ell)$ is the Fourier transform of the 
estimator kernel $\tilde W_{\alpha\beta}$:
\begin{eqnarray}
\label{Jell}
J(\ell) \equiv 
\left[\sum_{\Delta\theta_{\alpha\beta}} \tilde W_{\alpha\beta}\right]^{-1}
\sum_{\Delta\theta_{\alpha\beta}} \tilde W_{\alpha\beta}
{\,\rm exp} [-i \ell \cdot \Delta\theta_{\alpha\beta}] 
\end{eqnarray}

Among all the terms for $V (i)$, which correspond to those
considered by \cite{jt03}? They are the sum of terms in
$V^1 (i)$ and $V^2 (i)$ that consist
of the product of shape-noise and shot-noise (we will refer to
these loosely as shot-noise terms):
\begin{eqnarray}
&& V^{\rm \cite{jt03}} (i) \equiv \\ \nonumber 
&& [\int {d^2 \ell \over (2\pi)^2} P_{\rmg\kappa}^{i,1} (\ell) J(\ell)]^{-2}
{\sigma_\kappa^2 \over A_T \bar n^g_i \bar n_1^B} 
\int {d^2 \ell \over (2\pi)^2} {|J (\ell) |^2}
\\ \nonumber 
&& + 
[\int {d^2 \ell \over (2\pi)^2} P_{\rmg\kappa}^{i,2} (\ell) J(\ell)]^{-2}
{\sigma_\kappa^2 \over \bar A_T \bar n^g_i \bar n_2^B}
\int {d^2 \ell \over (2\pi)^2} {|J (\ell) |^2}
\end{eqnarray}
To see that this does correspond to what JT03 considered, note that
\cite{jt03} focused on the measurement of the real-space galaxy-convergence
correlation smoothed within some aperture (of, say, area $A_p$). This corresponds to a choice of $J(\ell)$
(or $\tilde W_{\alpha\beta}$ in eq.~[\ref{Pgshat}]) such
that $(2\pi)^{-2} \int d^2\ell |J (\ell) |^2 \sim 1/A_p$. 
Therefore, the above expression reduces to 
\begin{eqnarray}
&& V^{\rm JT03} (i) = {\sigma_\kappa^2 \over A_T A_P \bar n^g_i \bar n_1^B 
[P^{i,1}_{g\kappa}(A_P)]^2} \\ \nonumber
&& \quad \quad \quad \quad + {\sigma_\kappa^2 \over \bar A_T A_P \bar n^g_i \bar n_2^B
[P^{i,2}_{g\kappa}(A_P)]^2}
\end{eqnarray}
where we have abused the notation a little bit to denote the 
real-space galaxy-convergence correlation smoothed in an aperture
of area $A_P$ by $P^{i,1}_{g\kappa} (A_P)$. The above can be compared directly
with equation 12 of JT03. The dictionary for translating our
symbols to theirs is as follows: 
$\sigma_\kappa^2 \rightarrow \sigma_\epsilon^2/2$, 
$\bar n_1^B \rightarrow n_1$, $\bar n_2^B \rightarrow n_2$, 
$[P^{i,1}_{g\kappa}(A_P)]^2 \rightarrow \langle \gamma \rangle^2_{\ell 1}$,
$[P^{i,2}_{g\kappa}(A_P)]^2 \rightarrow \langle \gamma \rangle^2_{\ell 2}$,
$A_T \rightarrow A$, and $A_P \bar n^g_i \rightarrow f_\ell$. 
The last item requires a little explanation. JT03 defined
$f_\ell$ to be the fraction of the survey that is covered by
the apertures centered on foreground objects. This is equal to
$A_P \times (\bar n^g_i A_T) / A_T$, where $A_T$ is the total survey area.
With this, the correspondence with the expression of \cite{jt03} is manifest.

The expressions for statistical errors are actually simpler
in Fourier space
\ie suppose instead of measuring the galaxy- convergence
correlation smoothed in some aperture, one measures the
galaxy-convergence power spectrum at wavenumber $\ell$.
One can obtain the ratio $R^i$ for each $\ell$, and then
combine all these estimates of $R^i$ from each $\ell$ in
a minimum variance manner.
Note that while this is different from the procedure of \cite{jt03}, 
the procedure here will likely produce smaller errorbars on 
$R^i$, since it makes use of all information contained
in the modes instead of focusing on fluctuations at particular
scales.

Let us focus on a particular wavenumber (or band) $\ell$ for the moment.
Eq.s~[\ref{VR},\ref{VRs}] reduce to something quite
simple:
\begin{eqnarray}
\label{Viell}
V_\ell (i) = [P_{\rmg\kappa}^{i,1} (\ell)]^{-2} (P_{gg}^{i,i} (\ell) 
+ {1\over \bar n^g_i}) 
(P_{\kappa\kappa}^{1,1} (\ell) + {\sigma_\kappa^2 \over \bar n^B_1})
\\ \nonumber
+ [P_{\rmg\kappa}^{i,2} (\ell)]^{-2} (P_{gg}^{i,i} (\ell) 
+ {1\over \bar n^g_i}) 
(P_{\kappa\kappa}^{2,2} (\ell) + {\sigma_\kappa^2 \over \bar n^B_2})
\\ \nonumber
- 2 [P_{\rmg\kappa}^{i,1} (\ell)]^{-1}
[P_{\rmg\kappa}^{i,2} (\ell)]^{-1}
(P_{gg}^{i,i} (\ell) + {1\over \bar n^g_i}) P_{\kappa\kappa}^{1,2} (\ell)
\end{eqnarray}
where we have used $J (\ell) = |J (\ell)|^2 = 
(2\pi)^2 \delta^2 (\ell - \ell') / A_T$, and we have
introduced subscript $\ell$ to $V$ to emphasize this is the variance
of $R^i$ from Fourier bin $\ell$.

The Fourier analog of the approximation made by JT03 would be to retain only
the following terms in the variance:
\begin{eqnarray}
V_\ell^{\rm JT03} (i) = 
[P_{\rmg\kappa}^{i,1} (\ell)]^{-2} {1\over \bar n^g_i}{\sigma_\kappa^2 \over \bar n^B_1}
+ [P_{\rmg\kappa}^{i,2} (\ell)]^{-2} {1\over \bar n^g_i}
{\sigma_\kappa^2 \over \bar n^B_2}
\end{eqnarray}
This misses a number of terms compared to $V_\ell (i)$ in eq. [\ref{Viell}].
Each of the terms ignored by JT03 are of order unity. 
They can be thought of as sampling variance terms.
While there is some partial
cancellation among them, they do not cancel exactly and should be retained.

\cite{jt03} considered the constraint on dark energy from the ratio
$R^i$ for $i$ ranging over 10 different foreground redshift bins, 
ranging from $z = 0$ to $z = 1$, each with $\Delta z = 0.1$.
In addition to the diagonal variance considered above, 
there will in general be covariance between $R$ measured from
foreground bin $i$ and foreground bin $j$, 
which was not considered by \cite{jt03}:
\begin{eqnarray}
\label{Vijell}
&& V_\ell (i,j) \equiv \langle \delta \hat R^i \delta \hat R^j \rangle
/ (R^i R^j)  \\ \nonumber && = \delta_{ij} V_\ell (i) + 
[P_{\rmg\kappa}^{i,i1} (\ell)]^{-1}
[P_{\rmg\kappa}^{j,j1} (\ell)]^{-1} 
P_{\rmg\kappa}^{i,j1} P_{\rmg\kappa}^{j,i1} \\ \nonumber &&
+ [P_{\rmg\kappa}^{i,i2} (\ell)]^{-1}
[P_{\rmg\kappa}^{j,j2} (\ell)]^{-1} 
P_{\rmg\kappa}^{i,j2} P_{\rmg\kappa}^{j,i2} \\ \nonumber &&
- [P_{\rmg\kappa}^{i,i1} (\ell)]^{-1}
[P_{\rmg\kappa}^{j,j2} (\ell)]^{-1} 
P_{\rmg\kappa}^{i,j2} P_{\rmg\kappa}^{j,i1} \\ \nonumber &&
- [P_{\rmg\kappa}^{i,i2} (\ell)]^{-1}
[P_{\rmg\kappa}^{j,j1} (\ell)]^{-1} 
P_{\rmg\kappa}^{i,j1} P_{\rmg\kappa}^{j,i2}
\end{eqnarray}
Note the somewhat clumsy notation: 
instead of specifying the 2 background bins by just
$1$ and $2$, we now have to specify them by $i1$ and $i2$ which refers to
the 2 background bins that correspond to the $i$-th foreground bin,
and similarly for $j1$ and $j2$.
The covariance $V_\ell (i,j)$, when $i \ne j$, is non-vanishing
-- the positive and negative terms present do not exactly
cancel each other, and generically result in something
of same order of each of these terms, with perhaps some mild suppression.

\begin{figure}[tb]
\centerline{\epsfxsize=9cm\epsffile{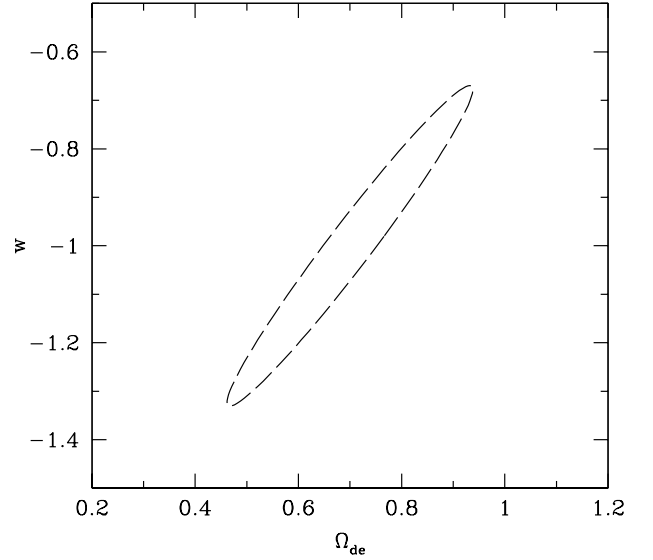}}
\caption{Dark energy constraints (1 $\sigma$) by adopting the linear scaling,
survey specifications and redshift binning of JT03.
No prior is placed on $\Omega_{\rm de}$ or $w$.
The fiducial model has $w = -1$ and $\Omega_{\rm de} = 0.7$;
$w'$ is fixed at $0$. 
}
\label{jtconstraints}
\end{figure}

Making use of $V_\ell (i,j)$ one can then work out the dark energy constraints
from the linear scaling of JT03.
Adopting the survey specifications and redshift-binning according to JT03,
we find constraints that are shown in Fig. \ref{jtconstraints}.
This can be compared against the smallest contour in Fig. 1 of JT03.
Hu \& Jain (2003) independently reached similar 
conclusions as in Fig. \ref{jtconstraints}.

In summary, it appears JT03 ignored certain contributions to
the variance (and covariance) of the ratio $R^i$.
They are primarily sampling variance terms.
These are automatically taken into account in our Fisher matrix analysis
in \S \ref{fisher}, which actually does not require an explicit computation
of all these variance terms. This should be contrasted with the Fisher matrix
calculation of BJ03: while we start with the galaxy-density and shear fields
as input Gaussian random data and compute constraints on parameters which enter
into the correlation matrix (eq. [\ref{BIGC}]), BJ03 started with
the quadratic estimates of lensing power spectra themselves as Gaussian
distributed input data. The latter approach requires explicit computation
of the variance and covariance of these quadratic estimates, and care
should be taken to include all contributions. It appears some of these
contributions were not included in the analysis of BJ03.
We have not, however, performed an analysis replicating the details of BJ03.

%From /merton1/lhui/Newfermi/Corr/Jun/TestJT03/ToLam/Bhuv_Vw100.ps
% or laptop: GeomSep5/FigAppendix/TestJT03/ToLam/Bhuv_Vw100.ps

\section*{Appendix B -- Non-flat Universe, Shear and Real Space Correlations}
\label{appendixB}

Our goal in this Appendix is to state our main results
in this paper for the more general case of a non-flat universe, 
for shear instead of convergence, and in real as well as Fourier space.
Some of the expressions have appeared in the literature. They are
given here for completeness.

Let us start with what is most commonly measured in galaxy-galaxy
lensing experiments, and relate it to the galaxy-convergence
power spectrum $P_{\rmg\kappa} (\ell)$ given 
in eq.~[\ref{Pgs}] (\cite{kaiser92}):
\begin{eqnarray}
\label{xiggamma}
\xi_{g\gamma^+} (\theta) = - \int {\ell d \ell \over 2\pi}
P_{\rmg\kappa} (\ell) J_2 (\ell \theta)
\end{eqnarray}
where $J_2$ is the second order Bessel function, 
\footnote{$J_n(y) = {1\over 2 \pi} \int_{-\pi}^\pi d\eta
{\,\rm cos} [y {\,\rm sin}\eta - n\eta]$}
and
$\xi_{g\gamma^+} (\theta)$ is the cross-correlation between
galaxies and tangential shear at separation $\theta$, a quantity
that is most commonly discussed in galaxy-galaxy lensing measurements. Alternatively, 
in a fixed coordinate system where $\gamma^1$ and $\gamma^2$
are the two components of shear, the 2 different
galaxy-shear power spectra $P_{g\gamma^1} (\ell)$ and
$P_{g\gamma^2} (\ell)$ are related to the galaxy-convergence
power spectrum $P_{\rmg\kappa}$ by:
\begin{eqnarray}
\label{Pggamma}
P_{g\gamma^1} (\ell) = {\,\rm cos} (2\phi_\ell) P_{\rmg\kappa} (\ell) \, , \,
P_{g\gamma^2} (\ell) = {\,\rm sin} (2\phi_\ell) P_{\rmg\kappa} (\ell)
\end{eqnarray}
where $\phi_\ell$ specifies the
orientation of the wavevector: $\ell {\,\rm cos}\phi_\ell$ is the
x-component while $\ell {\,\rm sin}\phi_\ell$ is the y-component.

Similarly, the two quantities that are commonly considered
in actual shear-shear correlation measurements are
related to the convergence power spectrum $P_{\kappa\kappa} (\ell)$
of eq.~[\ref{Pss}] by (\cite{kaiser92}):
\begin{eqnarray}
\label{xigammagamma}
\xi_{\gamma^+\gamma^+} (\theta) = 
{1\over 2} \int {\ell d\ell \over 2\pi}
P_{\kappa\kappa} (\ell) [J_0 (\ell\theta) + J_4 (\ell\theta)]
\\ \nonumber
\xi_{\gamma^\times \gamma^\times} (\theta) = 
{1\over 2} \int {\ell d\ell \over 2\pi}
P_{\kappa\kappa} (\ell) [J_0 (\ell\theta) - J_4 (\ell\theta)]
\end{eqnarray}
where $\gamma^+$ and $\gamma^\times$ are the tangential 
and ortho-tangential (or radial) shear defined with respect
to separation between two points of interest. 
Alternatively, the two different shear-shear power spectra
in a fixed coordinate system are related to the convergence
power spectrum by
\begin{eqnarray}
\label{Pgammagamma}
P_{\gamma^1\gamma^1} (\ell) =  {\,\rm cos}^2 (2 \phi_\ell)
P_{\kappa\kappa} (\ell) \\ \nonumber 
P_{\gamma^2\gamma^2} (\ell) =  {\, \rm sin}^2 (2\phi_\ell) 
P_{\kappa\kappa} (\ell)
\end{eqnarray}

The main results of this paper derive from writing $P_{\rmg\kappa} (\ell; f,b)$
and $P_{\kappa\kappa} (\ell; f,b)$, which are the galaxy-convergence
and convergence-convergence power spectra between a foreground bin $f$
and background bin $b$, in the form of eq.~[\ref{fullscaling}], and
noticing some of the terms are small, which leads to the
offset-linear scaling of eq.~[\ref{approxscaling}].

Let us first give the expressions for each term in 
eq.~[\ref{fullscaling}] (and eq.~[\ref{approxscaling}])
in the case of a non-flat universe. 
Then, we will discuss how similar expressions hold for shear measurements,
and in real space.

The non-flat space analogs of eq.s [\ref{chieff},\ref{FGAB}] are
\begin{eqnarray}
\label{FGHnonflat}
&& 
F(\ell; f)\equiv{3\Omega_{\rmm0}H_0^2 \over 2 c^2}\int{d\chi\over a}W_f(\chi)
{1\over r(\chi)} \\ \nonumber 
&& \quad \quad \quad \quad \quad \quad 
{\, \rm cs} (\chi) P_{g\delta} ({\ell \over r(\chi)}) \\ \nonumber
&& 
G(\ell; f) \equiv - {3\Omega_{\rmm0} H_0^2 \over 2 c^2} 
\int {d\chi\over a} W_f (\chi) 
{1 \over r(\chi)} \\ \nonumber 
&& \quad \quad \quad \quad \quad \quad 
{\, \rm si} (\chi)
P_{g\delta} ({\ell \over r(\chi)}) 
\\ \nonumber
&& I(\ell; f,b) \equiv - {3\Omega_{\rmm0} H_0^2 \over 2 c^2}
\int {d\chi\over a} W_f (\chi) 
\int d \chi' W_b (\chi') \\ \nonumber
&& \quad \quad \quad \quad
{r(\chi' - \chi) \over r(\chi') r(\chi)} P_{g\delta} 
({\ell \over r(\chi)}) 
\Theta(\chi-\chi')
\end{eqnarray}

\begin{eqnarray}
\label{chieffnonflat}
{1\over \chieff (b)} \equiv \int d\chi' W_b(\chi') {1\over 
{\rm ta} (\chi')} 
\end{eqnarray}

\begin{eqnarray}
\label{ABDnonflat}
&& A(\ell; f) \equiv \left({3\Omega_{\rmm0} H_0^2 \over 2 c^2}\right)^2
\int {d\chi''} W_f (\chi'') \\ \nonumber
&& \quad \quad \quad \int {d\chi \over a^2}
{r(\chi'' - \chi) \over r(\chi'')} 
{\rm cs} (\chi) P_{\delta\delta} ({\ell \over r(\chi)}) \Theta(\chi''-\chi) 
\\ \nonumber
&& B(\ell; f) \equiv - \left({3\Omega_{\rmm0} H_0^2 \over 2 c^2}\right)^2
\int {d\chi''} W_f (\chi'') \\ \nonumber 
&& \quad \quad \quad \int {d\chi \over a^2} {r(\chi'' - \chi) \over 
r(\chi'')} {\,\rm si} (\chi)
P_{\delta\delta} ({\ell \over r(\chi)})\,\Theta(\chi''-\chi)
\\ \nonumber
&& D(\ell; f,b) \equiv - \left({3\Omega_{\rmm0} H_0^2 \over 2 c^2}\right)^2
\int {d\chi''} W_f (\chi'') \int d \chi' W_b (\chi')
\\ \nonumber 
&& \quad \quad \int {d\chi \over a^2} {r(\chi' - \chi) \over r(\chi')}
{r(\chi'' - \chi) \over r(\chi'')} P_{\delta\delta} ({\ell\over r(\chi)}) \\
\nonumber 
&& \quad \quad \Theta(\chi-\chi')\,\Theta(\chi''-\chi)
\end{eqnarray}
where $r(\chi)$ is the comoving angular diameter distance
which is related to the comoving radial distance $\chi$
as follows:
$r(\chi) = K^{-1/2} {\,\rm sin} K^{1/2} \chi, 
(-K)^{-1/2} {\,\rm sinh} (-K)^{1/2} \chi, \chi$
for a closed, open and flat universe respectively, and
$K = - \Omega_k H_0^2/c^2$, where $\Omega_k$ is the curvature
in unit of the critical density. 
The quantities, ${\,\rm cs} (\chi)$, ${\,\rm si} (\chi)$, and
${\,\rm ta} (\chi)$ are defined as:
\begin{eqnarray}
{\,\rm cs} (\chi) = {\, \rm cos} K^{1\over 2} \chi, 
{\,\rm si} (\chi) = {\, \rm sin} K^{1\over 2} \chi, 
{\, \rm ta} (\chi) = {\, \rm tan} K^{1\over 2} \chi 
\end{eqnarray}
if $K > 0$, 
\begin{eqnarray}
{\,\rm cs} (\chi) = 1, {\,\rm si} (\chi) = \chi, {\,\rm ta} (\chi) = \chi
\end{eqnarray} 
if $K = 0$, and 
\begin{eqnarray}
{\,\rm cs} (\chi) = {\, \rm cosh} (-K)^{1\over 2} \chi, 
{\,\rm si} (\chi) = {\, \rm sinh} (-K)^{1\over 2} \chi, \\ \nonumber 
{\,\rm ta} (\chi) = {\, \rm tanh} (-K)^{1\over 2} \chi
\end{eqnarray} 
if $K < 0$. 

As before, the offset-linear scaling (eq.~[\ref{approxscaling}])
follows from eq.~[\ref{fullscaling}] 
by noticing that $D$ and $I$ are small provided
that $W_i$ and $W_j$ have little overlap, except that 
the relevant quantities
$A$, $B$, $D$, $F$, $G$ and $I$ are defined as above.
With the above expressions, one can in principle 
fit for $\Omega_k$ in addition
to the dark energy parameters in carrying out the exercise of
\S\ref{fisher}.

Lastly, it is trivial to generalize the offset-linear scaling of 
eq.~[\ref{approxscaling}] to galaxy-shear and shear-shear (instead
of galaxy-convergence and convergence-convergence as before) power spectra
by using eq.s~[\ref{Pggamma},\ref{Pgammagamma}]
\ie simply multiply eq.~[\ref{approxscaling}] by appropriate factors
of ${\, \rm sin} (2\phi_\ell)$ or ${\,\rm cos} (2\phi_\ell)$. 
Rewriting the scaling in real-space is no less difficult:
simply substitute eq.~[\ref{approxscaling}] into the expressions
for $\xi_{g\gamma^+} (\theta)$, $\xi_{\gamma^+\gamma^+}$
or $\xi_{\gamma^\times \gamma^\times}$ in eq.~[\ref{xiggamma}]
and (\ref{xigammagamma}). One can see that the scaling continues
to hold for real-space analogs of $A$, $B$, etc. 
In particular, eq.~[\ref{Pdiffratio}] holds for any of
these real space correlation functions \eg
\begin{eqnarray}
{\xi_{g\gamma^+} (\theta; f,b) - \xi_{g\gamma^+} (\theta; f,b') \over
\xi_{g\gamma^+} (\theta; f,b'') - \xi_{g\gamma^+} (\theta; f,b''')}
= {\chieff (b) ^{-1} - \chieff (b')^{-1}
\over \chieff (b'')^{-1} - \chieff (b''')^{-1}}
\end{eqnarray}
where $\xi_{g\gamma^+} (\theta; f,b)$ refers to the galaxy-tangential-shear
correlation between foreground redshift bin $f$ and background 
redshift bin $b$.

\end{document}